\documentclass[runningheads]{llncs}

\usepackage[T1]{fontenc}
\usepackage[utf8]{inputenc}
\usepackage{graphicx}
\usepackage{booktabs}
\usepackage{array}
\usepackage{amssymb}   
\usepackage{multirow}
\usepackage[table]{xcolor}
\usepackage{hyperref}

\newcommand{\yes}{\ensuremath{\bullet}}     
\newcommand{\pc}{\ensuremath{\odot}}         
\newcommand{\no}{\ensuremath{\circ}}         
\newcommand{\aff}{\ensuremath{\maltese}}     
\newcommand{\na}{\textendash}                

\begin{document}

\title{Recognition Without Mitigation:\\ Ethical Frameworks in Autonomous Offensive-LLM Agent Research}
\titlerunning{Ethical Frameworks in Autonomous Offensive-LLM Agent Research}

\author{Andreas Happe\orcidID{0009-0000-2484-0109} \and
Jürgen Cito\orcidID{0000-0001-8619-1271}}
\authorrunning{Happe and Cito}

\institute{TU Wien, Vienna, Austria} 

\maketitle

\begin{abstract}
Large language models have moved from advising on offensive security to autonomously
conducting it. A growing literature presents agents that execute reconnaissance,
exploitation, and privilege escalation against real or simulated targets. Such an
agent is a deployable, re-pointable capability that could be used by a malicious actor against a non-consenting third party. Papers that introduce these prototypes therefore carry an ethical burden, which top security
venues have begun to encode as hard policy in their 2026 ethics mandates. We
present a systematic audit of ethics reporting based on 54 papers describing autonomous offensive-LLM penetration-testing
prototypes (2023--2026), assembled from peer-reviewed venues as well as from pre-prints. We score each against an eleven-dimension instrument derived both top-down from the Menlo Report, and bottom-up from 2026 security venue ethics mandates.
Our central result is a \emph{recognition-without-mitigation} gap:
dual-use risk is reported as \emph{recognized} in 57\% of papers, but a \emph{concrete mitigation} is reported in only 15\%, roughly a 4:1 gap. Safeguards commonly protect the experiment, not the public. Guardrail bypasses are deployed by 17\% of papers but not disclosed to LLM providers. Measured against the new mandates, the corpus defines a
pre-regulation baseline in which current practice does not meet the substantive requirements. We argue this
audit is itself defensive intelligence on the offensive-agent ecosystem, and provide an ethics statement checklist for authors working on autonomous offensive agents.

\keywords{autonomous offensive security \and LLM agents \and research ethics \and dual-use
\and responsible disclosure}
\end{abstract}

\section{Introduction}
In under three years, large language models (LLMs) have crossed from \emph{advising} on
offensive security to \emph{conducting} it. A growing body of research presents systems in
which one or more LLMs autonomously select and execute
reconnaissance, exploitation, privilege escalation, and lateral movement against live or
simulated targets. Examples include early privilege-escalation agents~\cite{getpwnd}, pentest
assistants~\cite{pentestgpt}, web exploitation agents~\cite{fangoneday}, post-breach
attackers~\cite{autoattacker}, and attacks against live-network systems~\cite{incalmo,cochise}. These artifacts are not only
benchmarks that score a model. They are deployable, re-pointable offensive prototypes that could be used by a malicious actor against non-consenting third-parties. A capability of this kind places an ethical burden on the papers that introduce it. Unlike a measurement study or a classifier, an autonomous offensive agent can be aimed at a non-consented target by anyone who obtains it, and its potential harm grows automatically as foundation models improve. The security-research community has begun to encode this burden into hard policy: as of 2026, USENIX Security, IEEE S\&P, IEEE EuroS\&P, ACM CCS, NDSS, and ACM AsiaCCS state ethics requirements while reserving the right to reject papers that fail to comply with those~\cite{usenix26,sp26,eurosp26,ccs26,ndss26,asiaccs26}. Yet we have no empirical picture of how the offensive-agent literature actually reports ethics, and whether it merely \emph{names} the danger or actually \emph{contains} it.

To provide this picture, we assemble a corpus of \textbf{54} autonomous offensive LLM-driven penetration-testing prototypes (2023--2026) from two reproducible channels (a Scopus query and a forward-snowball of two seed papers) and audit each against an eleven-dimension instrument derived top-down from the Menlo Report's principles~\cite{menlo} and bottom-up from the 2026 conference mandates. Our approach follows the method of~\cite{pauley}, an audit of ethics frameworks in internet-measurement studies.

\subsection{Contributions and Findings}

\begin{enumerate}
\item \textbf{An empirical audit of papers on offensive use of autonomous agents} ($n=54$), with per-paper coding based on eleven ethical dimensions derived from 2026 conference mandates (Section~\ref{sec:instrument}).
\item Our central finding is a \textbf{recognition-without-mitigation gap}: dual-use is \emph{recognized} in 57\% of papers, but a \emph{concrete mitigation} is reported in only 15\%, roughly a 4:1 gap.

91\% of papers implement research-integrity controls, while only 15\% provide targeted counter-measures for defenders, thus \textbf{safeguards commonly protect the experiment, not the public}. LLM filtering and safeguards are frequently bypassed (17\% of papers), but \textbf{guardrail bypasses are not disclosed to LLM-providers} (Section~\ref{sec:reporting_practice}).
\item  Most of the corpus predates the 2026 mandates and so defines a \textbf{pre-regulation baseline}. Mapping the corpus against the requirements shows where the corpus fails the substantive requirements. We provide \textbf{a minimal ethics statement checklist} for offensive-agent papers (Section~\ref{sec:ethics_requirements}).
\end{enumerate}

\subsection{Research Questions}

\begin{description}
\item[RQ1.] To what extent do autonomous offensive-LLM prototype papers report ethical considerations across the eleven dimensions?
\item[RQ1a.] Does \emph{recognition} of dual-use risk translate into \emph{action} to prevent it?
\item[RQ2.] How does observed practice compare to venue ethics mandates?
\end{description}

\section{Background and Related Work}

\subsection{Autonomous Offensive-Security LLM Agents}
\label{sec:autonomous_research}

In recent years, LLM-driven agents that autonomously perform offensive actions against target systems have become routine research artifacts. For our study, we define an \textit{autonomous offensive-security LLM prototype} as a software system in which one or more LLMs drive a perception--action loop that selects and executes offensive actions (reconnaissance, exploitation, privilege escalation, lateral movement, exfiltration, or equivalent) against a target system.

\subsection{What an Ethics Statement Should Contain}
\label{sec:mandates}

The question of what an ethics statement should contain predates the use of LLMs for autonomous penetration-testing. We base our investigation on two different sources: the often cited Menlo Report~\cite{menlo} and ethical requirements and expectations by security venues as stated in 2026.

The Menlo Report supplies four principles: beneficence, respect for persons, justice, and respect for law and public interest. Based on the Menlo report, Partridge and Allman~\cite{partridge} apply those to derive what an
ethics section should contain in the field of internet measurements. This in turn has been taken up by Pauley and McDaniel~\cite{pauley} to audit concrete ethical statements in internet measurement papers.

Recently, the ``expectations'' question has hardened into policy. As shown in Table~\ref{tab:mandates}, requirements included in Call-for-Papers (CfP) are diverse. USENIX and S\&P unconditionally require an ethics section, while other conferences make it conditional on reported ethics. USENIX requires listing mitigations, or if no mitigation is offered, to provide a list of unmitigated risks. CCS makes ethics reporting conditional but, if applicable, defers to USENIX's requirements for details. USENIX expects a stakeholder-based ethics analysis including a weighting of harms and an explicitly stated decision to proceed with the research. NDSS echoes the impact analysis requirement through its ``Authors must proactively anticipate and address potential negative consequences of both conducting and publishing their research''. Multiple venues require coordinated vulnerability disclosure if a vulnerability is found. Similarly, an ethics board review or IRB is recommended if user data or users are concerned by the research. All but two (AsiaCCS and EuroS\&P) directly or indirectly reference the Menlo Report for guidance.

\begin{table}[t]
\caption{Per-venue ethics mandates for 2026. \checkmark\ required; (c) conditional/encouraged.}
\label{tab:mandates}
\centering\small
\setlength{\tabcolsep}{4pt}
\begin{tabular}{@{}lcccccc@{}}
\toprule
Mandated item                & USENIX     & S\&P    & EuroS\&P & CCS   & NDSS   & AsiaCCS \\
\midrule
1 Titled ethics section      & \checkmark & \checkmark & ---        & (c)       & (c)       & --- \\
2 Stakeholder analysis       & \checkmark & ---        & ---        & (c)       & ---       & --- \\
3 Benefit/harm analysis      & \checkmark & (c)        & \checkmark & (c)       & \checkmark & --- \\
4 Mitigations                & \checkmark & (c)        & ---        & (c)       & (c)       & --- \\
5 Dual-use / misuse          & (c)        & ---        & ---        & (c)       & ---       & --- \\
6 IRB / ethics-board discl.  & (c)        & (c)        & ---        & (c)       & (c)       & --- \\
7 Coordinated vuln. discl.   & (c)        & (c)        & ---        & (c)       & (c)       & --- \\
\midrule
Reject for non-compliance    & \checkmark & \checkmark & \checkmark & \checkmark & \checkmark & \checkmark \\
\bottomrule
\end{tabular}
\end{table}

All venues reserve the right to reject papers if ethics have not been sufficiently addressed. If requirements are not well defined, e.g., EuroS\&P and AsiaCCS provide one to two sentences, authors are left in the dark about what an ethics statement should actually contain.

\subsection{Empirical Audits of Ethics Statements}

We perform an empirical audit of ethics reporting across a corpus of published and pre-print publications. Its closest analogue is Pauley and McDaniel's audit of the ethical frameworks in internet-measurement studies~\cite{pauley}, which codes the ethical statements of measurement papers against a catalog of involved parties, data, and harms. It further compares the found ethics statements with ethics requirements of venue CfPs. In machine learning the same family appears as audits of NeurIPS broader-impact statements~\cite{nanayakkara,ashurst}.

\section{Methodology}\label{sec:method}

\subsection{Population and Operational Definition}

Based on the \textit{autonomous offensive agent} definition from Section~\ref{sec:autonomous_research}, we scope our study to papers where the primary contribution is the offensive agent system and empirical results are reported. Semi-autonomous (human-in-the-loop command gating, HITL) systems qualify because they still generate, propose, and execute offensive actions. We deliberately exclude benchmarks, capability evaluations, advisory chatbots, automatic-exploit-generation, and industry threat reports.

\subsection{Search Strategy and Corpus}
\label{search_strategy}

\begin{figure}[t]
\centering
\includegraphics[width=0.8\textwidth]{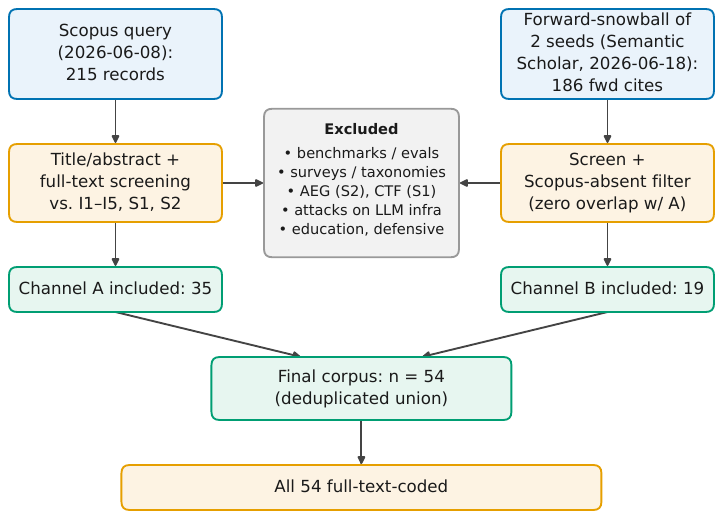}
\caption{Corpus selection (PRISMA-style): Channel~A (Scopus, 215$\rightarrow$35)
and Channel~B (reproducible forward-snowball of two seed papers, 186$\rightarrow$19). Final union $n=54$.}
\label{fig:prisma}
\end{figure}

We use a two-channel protocol with per-source reporting (Figure~\ref{fig:prisma}) yielding the final 54 papers shown in Table~\ref{tab:codingA}.

\paragraph{Channel A: Scopus.}

We used the following SCOPUS query to gather 215 potential papers on 2026-06-08:

\begin{quote}\scriptsize\ttfamily
TITLE-ABS-KEY(\\
\ ( "large language model*" OR "LLM" OR "LLM-based" OR "LLM-driven" OR "GPT-4" OR "GPT-3" OR "foundation model*" OR "generative AI" OR "language model agent*" )\\
\ AND\\
\ ( "penetration test*" OR "pentest*" OR "offensive security" OR "ethical hacking" OR \\
"privilege escalation" OR "vulnerability exploitation" OR "exploit generation" OR \\
"autonomous exploitation" OR "autonomous attack*" )\\
)\\
AND PUBYEAR > 2022\\
AND ( LIMIT-TO ( LANGUAGE , "English" ) )\\
AND ( LIMIT-TO ( DOCTYPE , "cp" ) OR LIMIT-TO ( DOCTYPE , "ar" ) OR LIMIT-TO ( DOCTYPE , "ch" ) )
\end{quote}

We used two scope rules to ``limit'' ourselves to penetration testing. Rule S1 excludes pure jeopardy-CTF solvers but keeps CTFs that target real networks/hosts. S2 excludes pure automatic-exploit-generation and payload generators with no pen-testing loop. Abstracts were subsequently screened against inclusion criteria I1--I5 (LLM-driven; executes offensive actions; the prototype is the contribution; empirical run; 2023--2026, English, peer-reviewed or arXiv). This finally yields \textbf{35 Channel-A inclusions}.

\paragraph{Channel B: reproducible forward-snowball.}
Because the field has a high number of preprint-only publications, Scopus under-indexes it. We recover the preprint
lineage with a reproducible supplement: a forward snowball of two seed papers (Channel A's two earliest and highly cited papers~\cite{getpwnd,pentestgpt}) via the Semantic Scholar citation graph (one hop, 2026-06-18). The 186 offensive-gated forward
citations were screened against the same I1--I5, S1/S2 criteria and \emph{restricted to papers
absent from the Channel-A Scopus result}. This yields \textbf{19 Channel-B inclusions}.

\subsection{The Eleven-Dimension Coding Instrument}
\label{sec:instrument}

\begin{table}[ht!]
\centering
\caption{The eleven-dimension instrument. This is a short representation of the codebook~\cite{zenodo}. Codes are classified into explicitly addressed (\yes), partially addressed (\pc), absent (\no), or not applicable (\na). For the two nominal dimensions D7b and D7c the levels are categories rather than a scale (see Sect.~\ref{sec:instrument}).}
\label{tab:instrument}

\begin{tabular}{p{12.3cm}}
\toprule

\begin{scriptsize}\textbf{D1 Ethics section:} does the paper contain (explicit) ethics statements? \yes\ if an explicit ethics section was contained in the paper, \pc\ if ethics statements were distributed throughout the work.\end{scriptsize}\\

\begin{scriptsize}\textbf{D2a Stakeholder \& Benefits:} are stakeholders and their benefits identified? \yes\ if more than one stakeholder including benefits was identified, \pc\ if a single stakeholder was identified.\end{scriptsize}\\

\begin{scriptsize}\textbf{D2b Impact \& Trade-Off}: are positive \emph{and} negative consequences discussed? Required two or more stakeholders. \pc\ if both negative and positive impacts were discussed, \yes\ if explicit weighting between harms was performed and a decision statement provided.\end{scriptsize}\\

\begin{scriptsize}\textbf{D3 Dual-Use Recognition}: is dual use explicitly mentioned? \pc\ if dual-use was mentioned, \yes\ if a concrete dual-use example or impact was highlighted.\end{scriptsize}\\

\midrule

\begin{scriptsize}\textbf{D4 Future-capability Acknowledgment:} does the paper analyze the impact of future models? \pc\ if only positive impact is discussed, \yes\ if the forecast also considers negative impact or harm.\end{scriptsize}\\

\midrule

\begin{scriptsize}\textbf{D5 Research Integrity:} are controls protecting the experiment and immediate target documented? These measures protect the research and lab environment, e.g., by keeping humans-in-the-loop or performing sandboxing. \yes\ if two orthogonal security measures were documented (Defense-in-Depth), \pc, if a single measure was documented.\end{scriptsize}\\

\midrule

\begin{scriptsize}\textbf{D6 IRB / ethics-review:} was involvement of an ethics board/IRB disclosed? Only applied, if human participants or third-party targets were involved. If the target was directly reached over the public internet, we included it due to the risk of accidents. \no\ if ethics review should be applicable but was not disclosed in the paper. \pc\ if any ethics review was mentioned, and \yes\ if an IRB review was explicitly performed.\end{scriptsize}\\

\midrule

\begin{scriptsize}\textbf{D7a Coordinated Disclosure:} were new vulnerabilities responsibly disclosed? \na\ if no new vulnerability was discovered, \no\ if no responsible disclosure was documented, \yes\ if responsible disclosure was documented.\end{scriptsize}\\

\begin{scriptsize}\textbf{D7b Guardrail Bypass Disclosure:} were bypassed safeguards disclosed? \na\ if no bypass was used, \no~if an intentional bypass is reported without provider disclosure, \pc~if a LLM was explicitly selected due to it having reduced guardrails, \yes~if a bypass was disclosed to the model provider.\end{scriptsize}\\

\begin{scriptsize}\textbf{D7c Release Conditions:} was source-code/prompts released to the public? \na\ if not mentioned within the paper, \no\ if source-code was published, \pc\ if no source-code was published but offensive prompts included in the paper, \yes\ if explicitly not published or gated-access for security reasons.\end{scriptsize}\\

\midrule

\begin{scriptsize}\textbf{D8 Recommendations for Defenders:} does the paper state how defenders can protect against malicious actors using this work against them? \yes\ if the recommendations were LLM-specific, \pc\ if generic recommendations were given.\end{scriptsize}\\

\bottomrule
\end{tabular}
\end{table}

We adopt the approach of Pauley and McDaniel~\cite{pauley} to derive the ethical dimensions which we will measure through our audit. We use top-level principles derived from the Menlo report and match them with the bottom-up requirements stated by recent security conferences (highlighted in Table~\ref{tab:crosswalk}). The resulting 11 dimensions are shown in Table~\ref{tab:instrument}. One dimension (D4) is not directly derivable from the venue mandates but was added as an LLM-specific extension. Four dimensions (D6, D7a, D7b, D7c) additionally carry a \emph{not-applicable} level (\na), used when the paper's target context, findings, or release status place the dimension out of scope. Both authors performed exploratory analysis to create the codebook~\cite{zenodo}. The corpus was coded by the first author, and corner cases were discussed with the second author. For quality control, two LLMs (Opus-4.8 and GPT-5.6-sol) were instructed to code the corpus using the same codebook. All conflicting codes were reviewed by both authors, and thus the final coding was derived.

  \begin{table}[t]
  \caption{Derivation crosswalk: each dimension mapped top-down to a Menlo principle and bottom-up
  to mandated item(s) of Table~\ref{tab:mandates}.}
  \label{tab:crosswalk}
  \centering\small
  \setlength{\tabcolsep}{4pt}
  \begin{tabular}{@{}llc@{}}
  \toprule
  Dimension & Menlo principle & Mandate \\
  \midrule
  D1 Ethics section          &  (container / all)            & 1    \\
  D2a Stakeholder \& benefit & Beneficence                   & 2    \\
  D2b Impact \& trade-off    & Beneficence, Justice          & 3    \\
  D3 Dual-use recognition    & Respect for Law \& Public     & 3,5  \\
  D4 Future-capability ack.  & Beneficence (over time)       & ---  \\
  D5 Research-integrity      & Beneficence; Respect for Law  & 4    \\
  D6 IRB / ethics-review    & Respect for Persons            & 6    \\
  D7a Coordinated disclosure & Respect for Law \& Public     & 7    \\
  D7b Guardrail bypass       & Respect for Law \& Public     & 7    \\
  D7c Release conditions     & Beneficence; Respect for Law  & 4    \\
  D8 Defender recommendations& Beneficence; Respect for Law  & 4    \\
  \bottomrule
  \end{tabular}
  \end{table}

\section{Results}

\begin{figure}[t]
\centering
\includegraphics[width=\textwidth]{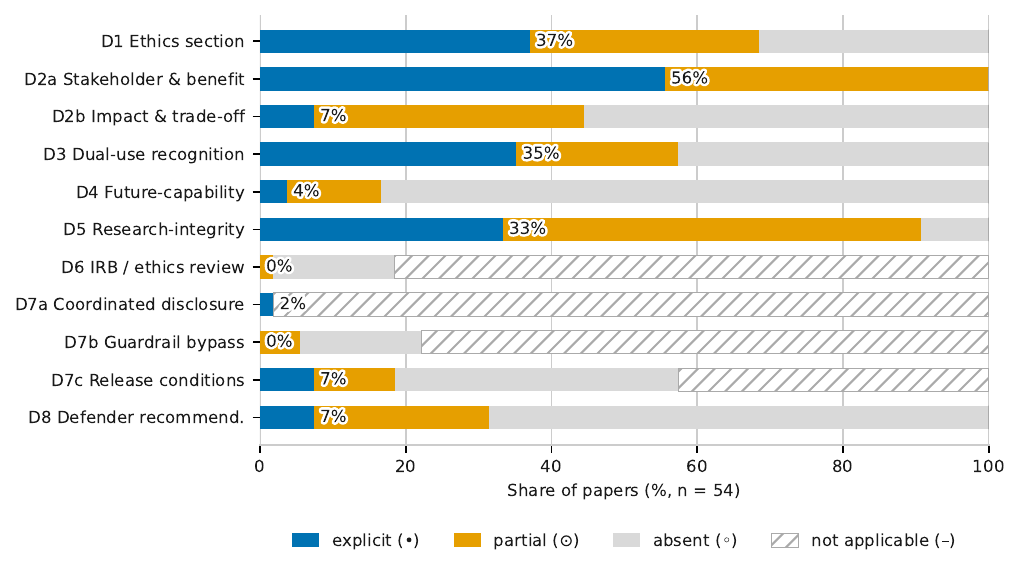}
\caption{Ethics reporting across the eleven dimensions, classified into explicitly addressed (\yes), partially addressed (\pc), absent (\no), or not applicable (\na). For the two nominal dimensions D7b and D7c the levels are categories rather than a scale (see Sect.~\ref{sec:instrument}).}
\label{fig:dimensions}
\end{figure}

The reviewed 54 papers and their codes are shown in Table~\ref{tab:codingA} (Appendix). Their summarization per ethical dimension is presented in Table~\ref{tab:proportions} and Figure~\ref{fig:dimensions}.

\subsection{Reporting Practice (RQ1)}
\label{sec:reporting_practice}

We differentiate between explicit (\yes) and partial (\pc) coverage of the ethical dimensions but commonly use the combined count (\yes\pc) to identify the papers' posture with regard to one dimension.

\paragraph{Overall Shape.} When looking at Figure~\ref{fig:dimensions}, almost all papers describe how they protect their lab environment from harm performed by the agent (D5, \yes\pc, 91\%), followed by discussion-side dimensions (D1--D3), i.e., topics that can be discussed as part of a paper (68\%, \yes\pc). Ethics reviews (D6) or vulnerability/bypass reporting (D7a/D7b) were commonly not applicable.

\paragraph{Discussion-side Dimensions (D1--D3).} 68\% of the papers contained ethics statements (D1, \yes\pc) of which 37\% were contained within a dedicated, separately-titled section (D1, \yes). All papers mentioned at least one stakeholder with their respective benefit provided by the work (D2a, \yes\pc). Benefits commonly given were ``scarcity of pen-testers''/``democratizing security'' (72\%, $n=39$) or ``preparing defenders'' (15\%, $n=8$). Only 56\% mentioned two or more stakeholders (D2a, \yes), indicating that 44\% were strongly focusing on a single stakeholder. This pattern continues with impact analysis, where positive and negative impacts were noted by 44\% of papers (D2b,\yes\pc\footnote{To achieve \yes\ on D2b, the paper must explicitly weigh the harm and benefits for stakeholders and explicitly state if the research should be released. Most analyzed papers did not include the explicit weighting or decision. As the papers were published, we counted that as implicit weighting and releasing, and thus use the aggregate number here.}). Notably more, 57\% (D3, \yes\pc), mentioned the dual-use and abuse potential of their prototypes.

\paragraph{Future-capability Awareness (D4).} 17\% of papers mentioned that future LLMs would impact the efficacy of their prototype. This was mostly seen as an improvement (D4, \yes\pc) with only two papers (4\%) explicitly stating that this also increases the abuse potential of their research (D4, \yes).

\begin{table}[ht]
\caption{Dimension proportions ($n=54$; \yes\,/\,\pc\,/\,\no\,/\,\na; raw counts in parentheses). Rows may deviate from 100\% by one point through rounding.}
\label{tab:proportions}
\centering\small
\setlength{\tabcolsep}{9pt}
\begin{tabular}{@{}lrrrr@{}}
\toprule
Dimension & \%\,\yes & \%\,\pc & \%\,\no & \%\,\na \\
\midrule
D1 Ethics section            & 37\% (20) & 31\% (17) & 31\% (17) & ---     \\
D2a Stakeholder \& benefit   & 56\% (30) & 44\% (24) & 0\% (0)   & ---     \\
D2b Impact \& trade-off      & 7\% (4)   & 37\% (20) & 56\% (30) & ---     \\
D3 Dual-use recognition      & 35\% (19) & 22\% (12) & 43\% (23) & ---     \\
D4 Future-capability ack.    & 4\% (2)   & 13\% (7)  & 83\% (45) & ---     \\
D5 Research-integrity        & 33\% (18) & 57\% (31) & 9\% (5)   & ---     \\
D6 IRB / ethics-review       & 0\% (0)   & 2\% (1)   & 17\% (9)  & 81\% (44) \\
D7a Coordinated disclosure   & 2\% (1)   & 0\% (0)   & 0\% (0)   & 98\% (53) \\
D7b Guardrail bypass         & 0\% (0)   & 6\% (3)   & 17\% (9)  & 78\% (42) \\
D7c Release conditions       & 7\% (4)   & 11\% (6)  & 39\% (21) & 43\% (23) \\
D8 Defender recommendations  & 7\% (4)   & 24\% (13) & 69\% (37) & ---     \\
\bottomrule
\end{tabular}
\end{table}

\paragraph{Experiment Environment Protection.} 91\% of the papers detailed how they protect their research from harm done by agents (D5, \yes\pc), e.g., sandboxing, virtualization, human confirmation or keeping humans-in-the-loop, or target limits within prompts. 33\% detail more than a single safety measure (D5, \yes). Please note, that these protections do not protect against dual-use of the prototype, but protect the experiment environment from it.

\paragraph{IRB/Ethics Review (D6).} We identified 9 papers (D6, \no) potentially benefiting from an ethics review. A single additional paper mentioned performing an ethical screening that indicated that no IRB involvement was necessary. While we agreed during our coding, we changed the code from \na\ to \pc\ to highlight that an ethics process was involved. Of the 9 papers marked \no, 4 were contacting a dedicated CTF or testing system over the public internet. We marked them \no~not for their danger against the testing host, but as accidents might impact unrelated third-parties due to public internet access.

\paragraph{Disclosure Policy (D7).} A single paper identified a traditional vulnerability and subsequently disclosed this to the vendor (D7a, \yes). 17\% of papers used a guardrail bypass (D7b, \no) to circumvent ethical filters, 6\% based their LLM-selection on models with reduced guardrails (D7b, \pc). No paper explicitly mentioned that a guardrail bypass was reported to LLM Providers.\footnote{Three papers (B1, B2, B4) contacted the LLM provider w.r.t. observed dangerous LLM capabilities but did not report bypassing guardrails or detecting novel vulnerabilities.} When it comes to releasing prototype code, 39\% of prototypes were publicly released (D7c, \no). Further 11\% (D7c, \pc) did not release the full source-code, but included offensive prompts in their publication. 7\% (D7c, \yes) explicitly stated that they will not release code due to security reasons, or that they only provide gated access. 43\% of papers did not state anything about the prototype availability. This amounted to 50\% of released prototypes (D7c, \pc\no), which attackers could directly abuse (dual-use).

\paragraph{Helping Defenders.} 31\% of publications mentioned countermeasures for defenders, i.e., providing defenders with means to protect themselves against malicious actors using the prototypes. Commonly very short, most were not specific to agentic attackers but instead described generic hardening measures (D8, \pc, 24\%). 7\% mentioned LLM-specific techniques, e.g., LLM tarpits or using LLM-specific patterns within IDS/IPS.

\paragraph{Retrieval Channel.} We compare publications within our two channels (Section~\ref{search_strategy}): Channel A only includes SCOPUS-indexed papers, which are peer-reviewed and published, while Channel B contains arXiv papers not included in Channel A. Discussion-side reporting (D1, D2b, D3) runs roughly $1.5\times$ as high in Channel B, and institutional-review disclosure, coordinated vulnerability disclosures, and conditioned releases are all Channel-B papers. Guardrail bypassing, in contrast, is channel-invariant (23\% vs.\ 21\%). Channel B also \emph{ships} more (D7c, \pc\no): 68\% of its papers release an artifact versus 40\% in Channel A. We include this information to highlight that Channel~B (pre-print) is not worse on any but one (D4) dimension we code, though no difference reaches significance at n=35/19.

\paragraph{The Impact of Releasing Open-Source Prototypes.}
\label{oss}

Table~\ref{tab:cm} compares countermeasures between papers that release their source-code or prompts (50\%, $n=27$) and papers that withheld them (50\%, $n=27$). The former have a higher awareness of dual-use (70\% vs 44\%).

When looking at papers that withhold sources or prompts, the act of withholding was the only concrete countermeasure detected. When including generic defensive recommendations, fewer papers describe countermeasures when compared to papers that release source code or prompts (26\% vs. 41\%).

\subsection{The Recognition--Mitigation Gap (RQ1a)}
\label{mitigation}

\begin{table}[t!]
\caption{Countermeasures in dependence of dual-use recognition (D3, \yes\pc) and release posture (released: D7c,\pc\no, withheld: D7c, \yes\na) Countermeasures (CM) is defined as papers disclosing vulnerabilities/bypasses, gating access, or providing LLM-specific recommendations for defenders (at least one \yes\ in (D7a, D7b, D7c, D8)). Generic countermeasures (GM) are defined as (D8, \pc). }
\label{tab:cm}
\centering\small
\setlength{\tabcolsep}{9pt}
\begin{tabular}{@{}lllrrr@{}}
\toprule
Population                  & Dual-Use & n         & CM        & CM+GM \\ 
\midrule

All                         & -             & 54        & 8 (15\%)      & 17 (33\%) \\
All                         & yes           & 31        & 7 (23\%)      & 16 (52\%) \\
Source/Prompts released     & -             & 27        & 4 (15\%)      & 11 (41\%) \\
Source/Prompts released     & yes           & 19        & 4 (21\%)      & 11 (58\%) \\
Source/Prompts withheld     & -             & 27        & 4 (15\%)      &  7 (26\%) \\
Source/Prompts withheld     & yes           & 12        & 3 (25\%)      &  5 (42\%) \\

\bottomrule
\end{tabular}
\end{table}

Table~\ref{tab:cm} summarizes the relationship between recognizing dual-use and implementing counter-measures. Dual-use is recognized in 57\% of papers (D3, \yes\pc). When looking at the whole population (``All''), countermeasures are reported in 15\% of papers, creating a \textbf{4:1 recognition--mitigation gap}. When including generic defense recommendations (D8, \pc), the ratio shrinks to 2:1 (33\%).

For the most charitable upper bound, we assume that the 43\% of papers not releasing their code\footnote{Without explicitly stating that this was done for security reasons.} do this as an abuse countermeasure, and also include generic countermeasures (D8,\pc). Further, we assume that authors who implement safeguards to protect their lab environment (D5) also implicitly see the dual-use potential of their work (n=52, D3+D5, \yes\pc). Even with those assumptions, only 38 papers (70\%) implement countermeasures against 52 (96\%) showing recognition, a ratio of 1.37:1.

Papers recognizing dual-use are more likely to offer at least generic defensive advice (52\% vs 9\%; Fisher exact, $p = 0.001$), but the difference disappears for concrete countermeasures (23\% vs 4\%, $p = 0.12$). Even among the 31 papers that recognize dual-use, 48\% report no countermeasure of any kind and 77\% report no concrete one.

\subsection{Compliance Against the 2026 Mandates (RQ2)}

\begin{table}[!t]
\caption{Observed practice against the seven mandated items of Table~\ref{tab:mandates} ($n=54$). ``Dim.''\ names the instrument dimension(s) each item is measured by.}
\label{tab:rq3}
\centering\small
\setlength{\tabcolsep}{4pt}
\begin{tabular}{lp{1cm}p{5.9cm}}
\toprule
Mandated item & Dim. & Observed practice \\
\midrule
1: Titled ethics section    & D1        & 37\% with dedicated ethics section (\yes), further 31\% carry embedded
                                            discussion (\pc). \\
2: Stakeholder analysis     & D2a       & 56\% name two or more stakeholders and benefits (\yes);
                                            31\% ($n=17$) include adversaries as an affected party.\\
3: Benefit/harm analysis    & D2b, D3  & harms \& benefits stated by 44\%, weighing and explicit decision documented by 7\%. 57\%  recognize dual-use potential.\\
4: Mitigations              & D5, D8, D7c & lab environment protected in 91\% of papers. Recommendations for defenders by 7--33\% (D8, either \yes or \yes\pc). 50\% of papers release source-code/prompts. \\
5: Dual-use / misuse        & D3        & 57\% recognize dual-use potential. \\
6: IRB / ethics-board discl.& D6        & no IRB involvement documented for 9 potentially applicable cases. \\
7: Coordinated vuln.\ discl.& D7a, D7b  & The single paper identifying new vulnerabilities performed responsible disclosure. No Guardrail bypasses (17\%, $n=9$) were reported to LLM providers.\\
\bottomrule
\end{tabular}
\end{table}

Table~\ref{tab:rq3} maps each mandated item of the per-item checklist (Table~\ref{tab:mandates}) to the practice we observe. Had the 54 papers been submitted under the 2026 rules, most of them would not comply to the ethics mandates. Only 37\% carry the titled ethics section two venues have as hard requirement (USENIX Security and IEEE S\&P). While stakeholders were identified by all papers, they often lack the thoroughness USENIX required. This can also be seen with the lack of a weighting of harms and few papers explicitly stating why they go forward with their research. Mitigations are often neglected and no IRB involvement disclosed, even if warranted. Coordinated vulnerability disclosure was performed for traditional vulnerabilities, but not for safeguard bypasses.

\section{Discussion}

\subsection{Recognition Without Mitigation}

An ethics \emph{statement} is not an ethics \emph{practice}. The identified \emph{recognition--mitigation gap} (Section~\ref{mitigation}) echoes findings of one of the papers inspiring this work~\cite{ashurst}: a norm's existence does not by itself produce harm-handling. This has an implication for venue requirements: they should focus on \textit{misuse containment} and \textit{documenting countermeasures}, not \textit{discussion}. Instead of asking ``who could be harmed?'', the question should be ``what concrete steps prevent it, and what residual harm is accepted?''

While our corpus was constructed with papers focusing on offensive penetration testing, we also investigated if papers included recommendations for defenders as potential mitigation. Only 7\% of papers documented concrete countermeasures. When including generic recommendations, e.g., system hardening or attack surface reduction, the rate rose to 31\%. This could be partially explained by only a single paper reporting new vulnerabilities. This indicates that the other papers were applying well-known attacks and techniques, against which generic countermeasures would help protect.

\subsection{Integrity vs. Security Controls}

While 91\% of papers protect their lab environment from accidents triggered by agents, only 15\% provide concrete counter-measures on how to protect against malicious actors using their prototypes against unconsenting third parties. One way of preventing abuse is to not release the prototype's source code. 50\% of papers were doing this, although only 7\% explicitly reported this as a security measure.

Papers that publicly release prototypes had a higher acknowledged rate of dual-use and proposed more countermeasures (Section~\ref{oss}) compared to non-releasing papers. There is friction between academic reproducibility, e.g., releasing prototypes, and trying to prevent harm through gated access~\cite{happe2026ethics}.

\subsection{Model Guardrail Bypass-Reporting Absent}

One paper discovered novel vulnerabilities and disclosed those to the respective vendors. Three papers notified LLM providers about dangerous capabilities observed during testing. This contrasts with the 17\% of papers (D7b, \no) that bypassed LLM guardrails and did not report them back to LLM providers. Prompt-level refusal suppression reaches back to the field's two seed papers. This indicates that guardrail bypasses are seen as a ``byproduct'', are normalized, and thus not reported back to the LLM providers as something noteworthy. We observed this behavior in both retrieval channels at nearly identical rates (21--23\%). For defense, this implies that provider-side alignment cannot be assumed to hold against a determined, research-grade adversary and thus should not be relied on.

\subsection{A Pre-Regulation Baseline: Will the Mandates Close the Gap?}

Because most of the corpus predates the 2026 mandates, it is a \emph{pre-regulation baseline}. The baseline frames a skeptical prediction: most mandates can be fulfilled by documenting stakeholders and potential harm, fitting our recognition-not-mitigation finding. If a prototype targets an isolated air-gapped lab environment without human participants, an ethical review is typically not triggered. As observed, prototypes predominantly industrialize existing attacks (and do not discover new vulnerabilities or attack classes); disclosure is not applicable. While the increased awareness of ethics is beneficial, closing the gap needs mandates to ask for D7c (disclosure) and D8-style (helping defenders) content explicitly.

\section{Ethics Requirements/Checklist for Offensive Research}
\label{sec:ethics_requirements}

Based on the 2026 venue mandates (Section~\ref{sec:mandates}), we derived an ethics requirements checklist. The ethics mandates can be classified into three groups. 

The first group is focused on documenting and discussing the paper's ethical context. A publication should (1) have a \emph{dedicated titled ethics section}. Within the work it (2) should \emph{identify multiple stakeholders} and how they are \emph{impacted by the work}. Based on the stakeholders and impact, the work should (3) perform a \emph{benefit/harm analysis weighting consequences} and explicitly state why the research was continued and published. If negative impact was documented, it (4) should either be \emph{mitigated}, or un-mitigated harms should be documented. This should include both harms that might occur during experiment execution, i.e, within the testing or lab environment, as well as harm that occurs when running the prototype in the wild. Given that publications will be about offensive tooling, (5) \emph{dual-use potential}, i.e., abuse through malicious actors, should be identified and concrete scenarios and their impact included in the analysis.

We further introduce a new requirement, tailored towards LLMs: the ethics statement should document how \emph{future model improvements} could change their impact analysis. 

The second group is dependent on the paper's methodology and findings. If humans are involved during the evaluation, or if third-party systems are targeted, (6) an \emph{IRB/ethics board should be involved}. We extend this by recommending to also involve the ethics board if communication with the target system occurs through the public internet as this implies that the offensive agent has public internet access and thus accidents involving unconsenting third-parties can occur. If humans or third-parties are involved, an IRB/ethics-board should be involved even if their consent was obtained. If vulnerabilities are discovered during the research, (7) \emph{responsible vulnerability disclosure} with the respective vendors should be performed. In addition to vulnerability disclosure, also model \emph{safeguard bypasses should be disclosed} to the respective LLM provider.

The third group is about preventing abuse through malicious parties. As the research details offensive tooling with dual-use potential, the paper should \emph{include recommendations how defenders can protect} against those tools.

\section{Defensive Implications}\label{sec:defensive}

The audit's primary value for autonomous cyber-defense is as \emph{threat intelligence on the offensive-agent supply chain} and what the offensive-research community releases and discloses. Three consequences follow:

\paragraph{1. Calibrate the threat model to unaligned agents.}

17\% of prototypes report bypassed model safety controls while 6\% choose models with reduced safeguards. A defender who assumes the adversary's LLM will refuse harmful instructions is calibrating to the wrong threat. At least 21 prototypes (39\% of corpus) released their source-code publicly and thus are offensive artifacts freely obtainable by any malicious actor.

\paragraph{2. Model providers are a first-class defensive stakeholder.}

If a prototype is released to the public, adversaries can remove or bypass D5-style safety guardrails and thus use a prototype as offensive tooling. One potential defense is LLM providers, which can implement structured access~\cite{shevlane2022structuredaccessemergingparadigm} and screen requests for malicious content. Because this inherently differentiates between users who can access cybersecurity capabilities and those who cannot, this security measure raises questions of fairness and policing.

\paragraph{3. Establish a common AI guardrail-bypass disclosure practice.}

To aid model providers in implementing safeguards, a new common disclosure practice and increased disclosure awareness are needed. Model providers should be given enough time to implement safeguards before new attacks or usage patterns become common knowledge.

\section{Conclusion}

We presented a systematic audit of ethics reporting in autonomous offensive LLM-powered penetration-testing prototypes. Papers universally make the case \emph{for} their work but rarely report how potential harm could be \emph{mitigated}: dual-use is recognized roughly four times more often than a mitigation is reported. Safeguards commonly protect the experiment, not the public. Model Guardrails are frequently bypassed (17\%), but not reported back to the LLM providers. Our surveyed papers often do not meet venues' ethics mandates (which predate these mandates). Because the mandates check for documentation, not mitigation, the recognition--mitigation gap may survive them. For autonomous cyber-defense, the audit doubles as threat intelligence. In addition, we offer a minimal ethics checklist for researchers.

\subsubsection*{Ethical Considerations.}

This is a meta-research study of \emph{published} papers; it involves no human subjects, no live systems, and no new offensive capability. We pass no judgment on conduct, only on what is visible in the papers, and frame ``absent'' as a reporting gap, not an accusation. We excluded no prototypes on the basis of authorship.

\subsubsection*{AI Disclosure.} The authors used Anthropic Claude Opus 4.8 to support corpus screening and revising texts. As part of quality-assurance, Claude Opus 4.8 and GPT-5.6-sol were used to code papers based on the existing codebook. All disagreements were reviewed by the authors. All research design, source interpretation, and conclusions are the authors' own.

\bibliographystyle{splncs04}
\bibliography{paper_redone}

\appendix
\section{Appendix}

\begin{table}[h]
\caption{Channel~A (Scopus, n=35) and Channel~B (Arxiv Snowball, n=19)}
\label{tab:codingA}
\centering\scriptsize
\setlength{\tabcolsep}{2.4pt}
\begin{tabular}{@{}rlcccccccccccc@{}}
\toprule
\# & System & Venue & D1 & D2a & D2b & D3 & D4 & D5 & D6 & D7a & D7b & D7c & D8 \\
\midrule
A1   & Getting pwn'd \aff\ (seed)~\cite{getpwnd} & ESEC/FSE'23 & \yes & \yes & \pc & \yes & \no & \yes & \no & \na & \no & \no & \no \\
A2   & PentestGPT (seed)~\cite{pentestgpt} & USENIX'24 & \pc & \yes & \no & \pc & \no & \yes & \no & \na & \no & \no & \no \\
A3   & PTGroup~\cite{ptgroup} & ICIC'24 & \no & \pc & \no & \no & \pc & \pc & \na & \na & \na & \na & \no \\
A4   & PENTEST-AI~\cite{pentestai} & IEEE CSR'24 & \no & \yes & \no & \no & \no & \no & \no & \na & \na & \na & \no \\
A5   & Autonomous Cyberattack~\cite{autoncyberattack} & IEEE CSR'24 & \pc & \yes & \no & \pc & \pc & \pc & \na & \na & \pc & \na & \no \\
A6   & AutoPen~\cite{autopen} & CSAE'25 & \no & \pc & \no & \no & \no & \pc & \na & \na & \no & \na & \no \\
A7   & Perses~\cite{perses} & AsiaCCS'25 & \pc & \pc & \pc & \yes & \no & \pc & \na & \na & \pc & \pc & \yes \\
A8   & CurriculumPT~\cite{curriculumpt} & App.Sci.'25 & \yes & \pc & \no & \pc & \no & \yes & \na & \na & \na & \na & \no \\
A9   & Auto Pentest GenAI~\cite{autopentestgenai} & ICGS3'25 & \yes & \yes & \no & \yes & \no & \pc & \na & \na & \na & \na & \no \\
A10  & VSFTPD~\cite{vsftpd} & ICTCS'25 & \yes & \yes & \pc & \pc & \no & \pc & \na & \na & \na & \no & \pc \\
A11  & RefPentester~\cite{refpentester} & IEEE PST'25 & \pc & \pc & \no & \no & \no & \yes & \na & \na & \na & \na & \no \\
A12  & AutoPentestAL~\cite{autopentestal} & TrustCom'25 & \no & \pc & \no & \no & \pc & \pc & \na & \na & \na & \na & \no \\
A13  & Controller~\cite{controllerpt} & TrustCom'25 & \no & \pc & \no & \no & \pc & \pc & \na & \na & \no & \no & \no \\
A14  & PExpAgent (C2)~\cite{pexpagent} & IEEE NaNA'25 & \pc & \pc & \no & \no & \no & \yes & \na & \na & \na & \na & \no \\
A15  & Agentic SSH~\cite{agenticssh} & Cyber-AI'25 & \pc & \pc & \no & \pc & \no & \yes & \no & \na & \na & \pc & \pc \\
A16  & Automated Pentest Tool~\cite{autopentesttool} & WorldSUAS'25 & \no & \yes & \no & \no & \no & \no & \na & \na & \na & \na & \pc \\
A17  & Agentic RAG~\cite{agenticrag} & SPCNC'25 & \no & \pc & \no & \no & \no & \pc & \na & \na & \na & \na & \no \\
A18  & LIMA~\cite{lima} & FMLDS'25 & \no & \yes & \no & \pc & \no & \yes & \na & \na & \na & \no & \pc \\
A19  & LLMs as Hackers \aff~\cite{llmhackers} & EMSE'26 & \yes & \yes & \yes & \yes & \no & \pc & \na & \na & \na & \no & \pc \\
A20  & AutoSecAgent~\cite{autosecagent} & J.Supercomp.'26 & \yes & \pc & \no & \yes & \no & \yes & \no & \na & \na & \na & \pc \\
A21  & PTFusion~\cite{ptfusion} & Inf.Fusion'26 & \no & \pc & \no & \no & \no & \pc & \na & \na & \na & \na & \no \\
A22  & SEPTA~\cite{septa} & ARTIIS'26 & \no & \pc & \no & \no & \no & \pc & \na & \na & \na & \pc & \no \\
A23  & PenAgent~\cite{penagent} & ISCAIT'26 & \no & \pc & \no & \no & \no & \pc & \na & \na & \na & \na & \no \\
A24  & ShellGPT / IACIS~\cite{shellgpt} & IACIS'25 & \pc & \pc & \pc & \yes & \no & \no & \no & \na & \na & \no & \no \\
A25  & Pentest LLM+RL~\cite{llmrlpentest} & IWANN'26 & \pc & \pc & \no & \pc & \no & \no & \na & \na & \na & \na & \no \\
A26  & SatGuard~\cite{satguard} & Aerospace'25 & \pc & \yes & \pc & \pc & \no & \yes & \na & \na & \no & \na & \no \\
A27  & Hierarchical PPO+LLM~\cite{hierppo} & ICIC'25 & \no & \pc & \no & \no & \no & \pc & \na & \na & \na & \na & \no \\
A28  & GPT-4o CTF (SSH)~\cite{gpt4octf} & IEEE CARS'25 & \pc & \yes & \pc & \yes & \no & \no & \no & \na & \na & \pc & \pc \\
A29  & Cochise \aff~\cite{cochise} & TOSEM'26 & \yes & \yes & \yes & \yes & \no & \yes & \na & \na & \na & \no & \yes \\
A30  & COAPT~\cite{coapt} & BDMA'26 & \pc & \yes & \no & \pc & \no & \pc & \na & \na & \na & \na & \pc \\
A31  & Trinity~\cite{trinity} & IEEE ICTMIM'26 & \pc & \yes & \no & \no & \no & \yes & \na & \na & \na & \no & \no \\
A32  & PenTest2.0~\cite{pentest2} & AINA'26 & \pc & \yes & \pc & \no & \pc & \yes & \na & \na & \na & \na & \no \\
A33  & CyberPentest~\cite{cyberpentest} & IEEE ACIT'25 & \no & \yes & \no & \no & \no & \yes & \na & \na & \na & \na & \no \\
A34  & AI-Assisted Local LLM+RAG~\cite{aiassistedlocal} & IEEE ACIT'25 & \no & \yes & \no & \no & \pc & \yes & \na & \na & \na & \na & \no \\
A35  & RedTeamLLM~\cite{redteamllm} & IJCAI'25 & \pc & \pc & \pc & \yes & \yes & \yes & \na & \na & \no & \no & \no \\
\midrule
B1   & Fang one-day~\cite{fangoneday} & 2404.08144 & \yes & \pc & \pc & \yes & \no & \pc & \na & \na & \na & \yes & \pc \\
B2   & Fang HPTSA (zero-day)~\cite{fanghptsa} & 2406.01637 & \yes & \pc & \pc & \yes & \no & \pc & \na & \na & \na & \yes & \pc \\
B3   & AutoAttacker~\cite{autoattacker} & 2403.01038 & \yes & \yes & \pc & \yes & \yes & \pc & \pc & \na & \no & \pc & \yes \\
B4   & Incalmo~\cite{incalmo} & 2501.16466 & \yes & \yes & \yes & \yes & \pc & \pc & \na & \na & \na & \no & \pc \\
B5   & VulnBot~\cite{vulnbot} & 2501.13411 & \no & \pc & \no & \no & \no & \pc & \na & \na & \no & \no & \no \\
B6   & RapidPen~\cite{rapidpen} & 2502.16730 & \yes & \yes & \pc & \yes & \no & \pc & \na & \na & \na & \na & \no \\
B7   & BreachSeek~\cite{breachseek} & 2409.03789 & \no & \yes & \no & \no & \no & \pc & \na & \na & \na & \no & \no \\
B8   & APT-Agent~\cite{aptagent} & 2605.24949 & \yes & \yes & \pc & \pc & \no & \pc & \na & \na & \na & \no & \no \\
B9   & AWE~\cite{awe} & 2603.00960 & \no & \yes & \no & \no & \no & \pc & \na & \na & \na & \no & \no \\
B10  & Environment-Grounded MAW~\cite{envgroundedmaw} & 2603.24221 & \pc & \pc & \no & \no & \no & \yes & \na & \na & \na & \yes & \no \\
B11  & PenForge~\cite{penforge} & 2601.06910 & \pc & \pc & \pc & \no & \no & \pc & \na & \na & \na & \no & \no \\
B12  & WiFiPenTester~\cite{wifipentester} & 2601.23092 & \yes & \yes & \pc & \yes & \no & \yes & \no & \na & \na & \yes & \pc \\
B13  & Cybersecurity AI --- Robots~\cite{cybersecairobots} & 2603.08665 & \yes & \yes & \pc & \yes & \no & \pc & \no & \yes & \na & \no & \yes \\
B14  & Enhancing Linux PrivEsc \aff~\cite{enhancingprivesc} & 2604.27143 & \yes & \yes & \yes & \pc & \no & \pc & \na & \na & \pc & \no & \pc \\
B15  & Post-Training PrivEsc \aff~\cite{posttraining} & 2603.17673 & \yes & \pc & \no & \yes & \no & \pc & \na & \na & \na & \no & \no \\
B16  & Guided Reasoning / STT~\cite{guidedreasoning} & 2509.07939 & \yes & \yes & \pc & \no & \no & \pc & \na & \na & \na & \no & \no \\
B17  & Hacking, The Lazy Way (Goyal)~\cite{goyal} & 2409.09493 & \yes & \yes & \pc & \yes & \no & \yes & \na & \na & \no & \na & \no \\
B18  & AutoEG~\cite{autoeg} & 2604.00704 & \yes & \yes & \pc & \pc & \no & \pc & \na & \na & \na & \pc & \no \\
B19  & APIOT~\cite{apiot} & 2605.02346 & \pc & \yes & \pc & \yes & \no & \pc & \na & \na & \na & \no & \no \\
\bottomrule
\end{tabular}\\[2pt]

{\scriptsize Legend: \yes\ explicit $\cdot$ \pc\ partial $\cdot$ \no\ absent $\cdot$ \na\ not applicable.
For the nominal dimensions the levels read: D7b \yes\ bypass disclosed to provider, \no\ intentional bypass undisclosed, \pc\ reduced-safeguard model chosen; D7c \yes\ conditioned access, \pc\ no code but full offensive prompts printed, \no\ fully open release, \na\ neither released. Channel~A $=$ Scopus (variant D); Channel~B $=$ reproducible forward-snowball of the two seed papers (Scopus-absent). \aff\ $=$ author-affiliated.}
\end{table}

\end{document}